\documentclass[acmsmall]{acmart}
\usepackage{amsmath,amsfonts,graphicx,url}
\usepackage{booktabs}
\usepackage{float}
\usepackage{xcolor}

\setcopyright{none}
\renewcommand\footnotetextcopyrightpermission[1]{}

\begin{document}
\title{Infrastructure First: Enabling Embodied AI for Science in the Global South}

\author{Shaoshan Liu}
\affiliation{
  \institution{Shenzhen Institute of Artificial Intelligence and Robotics for Society (AIRS)}
  \country{China}}
\email{shaoshanliu@cuhk.edu.cn}

\author{Jie Tang}
\affiliation{
  \institution{South China University of Technology}
  \country{China}}

\author{Marwa S. Hassan}
\affiliation{
  \institution{National Research Centre (NRC)}
  \country{Egypt}}

\author{Mohamed H. Sharkawy}
\affiliation{
  \institution{British University in Egypt (BUE)}
  \country{Egypt}}

\author{Moustafa M. G. Fouda}
\affiliation{
  \institution{National Research Centre (NRC)}
  \country{Egypt}}

\author{Tiewei Shang}
\affiliation{
  \institution{National Institute of Clean-and-Low-Carbon Energy}
  \country{China}}

\author{Zixin Wang}
\affiliation{
  \institution{National Institute of Clean-and-Low-Carbon Energy}
  \country{China}}

\begin{abstract}
\textbf{\textit{Abstract:}} Embodied AI for Science (EAI4S) brings intelligence into the laboratory by uniting perception, reasoning, and robotic action to autonomously run experiments in the physical world. For the Global South, this shift is not about adopting advanced automation for its own sake, but about overcoming a fundamental capacity constraint: too few hands to run too many experiments. By enabling continuous, reliable experimentation under limits of manpower, power, and connectivity, EAI4S turns automation from a luxury into essential scientific infrastructure. The main obstacle, however, is not algorithmic capability. It is infrastructure. Open-source AI and foundation models have narrowed the knowledge gap, but EAI4S depends on dependable edge compute, energy-efficient hardware, modular robotic systems, localized data pipelines, and open standards. Without these foundations, even the most capable models remain trapped in well-resourced laboratories. This article argues for an infrastructure-first approach to EAI4S and outlines the practical requirements for deploying embodied intelligence at scale, offering a concrete pathway for Global South institutions to translate AI advances into sustained scientific capacity and competitive research output.
\end{abstract}

\maketitle

\section{Introduction}
\label{sec:intro}

Scientific discovery is inherently physical. Samples must be prepared, instruments operated, experiments executed, and results verified in the real world. While artificial intelligence has transformed data analysis and modeling, much of the experimental process itself remains manual, slow, and labor-intensive. This mismatch limits how quickly laboratories can iterate and how broadly scientific capability can scale.

Embodied AI offers a way to close this gap. By integrating perception, reasoning, and action into physical systems, enabling machines to sense their environments, learn from interaction, and autonomously carry out tasks rather than merely provide software assistance ~\cite{fan2025putting}. Extending this paradigm to laboratories gives rise to Embodied AI for Science (EAI4S): intelligent agents that can observe experimental states, operate instruments, execute procedures, and adapt decisions directly within the scientific workflow ~\cite{eai4s}.

The foundations of this idea have steadily matured over two decades. Early robotic scientists demonstrated that machines could autonomously generate hypotheses and perform experiments end-to-end ~\cite{king2004functional}, and later work formalized the automation of the scientific method itself ~\cite{king2009automation}. More recent platforms combine robotics, mobility, and large language models to orchestrate complex laboratory operations over extended periods ~\cite{burger2020mobile}. Together, these developments signal a shift from AI as an analytical assistant to AI as an experimental actor.

This shift has profound implications for scientific productivity. Embodied systems can operate continuously, execute repetitive procedures reliably, and iterate experiments far faster than manual workflows. They reduce the marginal cost of experimentation, increase throughput, and improve reproducibility by standardizing execution. In effect, intelligence becomes a multiplier of laboratory capacity, allowing small teams to accomplish what previously required large, specialized staffs.

Such capability is especially consequential for the Global South. Worldwide scientific output remains highly concentrated in high-income regions, reflecting unequal access to experimental infrastructure rather than unequal access to ideas ~\cite{output}. For institutions facing limited personnel and resources, the ability to automate experimentation is not a luxury but a practical means to expand research capacity and participate more fully in frontier discovery. By lowering the cost and expertise required for high-quality experimentation, EAI4S presents a concrete opportunity to narrow the scientific gap between developing and advanced economies.

To ground this argument, we first examine how EAI4S quantitatively changes the productivity of scientific experimentation, before turning to a concrete laboratory case study and its implications for Global South research systems.

\section{How EAI4S Multiplies Scientific Productivity}
\label{sec:productivity}

We begin by reviewing empirical evidence across multiple scientific domains showing that EAI4S systematically increases experimental throughput beyond what is feasible in human-centered workflows.

EAI4S multiplies scientific productivity by increasing the duty cycle and throughput of physical experimentation while reducing the human time required per experimental cycle. In biology, the Robot Scientist Adam demonstrated automated, hypothesis-driven experimentation at a scale that is difficult to match manually, with reports that it can initiate up to 1,000 experiments in a day and run them with minimal human intervention ~\cite{Sparkes_2010}.

In chemistry and catalysis, a mobile robotic chemist operated autonomously for eight days and executed 688 experiments in a ten-variable design space, a sustained pace enabled by automated handling, measurement, and closed-loop decision-making rather than continuous staffing ~\cite{burger2020mobile}.

In materials science, autonomous laboratories have demonstrated rapid, minimally supervised synthesis and characterization workflows; for example, an autonomous materials synthesis platform reported realizing target materials at a rate of more than two additional materials per day with minimal human intervention ~\cite{Szymanski_2023}.

Complementing these platforms, high-throughput self-driving experimentation in device fabrication has reported automated film formation enabling up to 6,048 films per day, illustrating how embodied automation can compress large experimental design spaces into tractable campaigns ~\cite{Langner_2020}.

As a practical human baseline, conventional laboratory workflows are typically constrained to only a few experiments per operator per day, and in many organic synthesis settings rarely exceed approximately ten reactions per day, once synthesis, processing, analysis, and instrument access are accounted for. These limits arise from inherently serial manual operations and restricted duty cycles in human-centered experimentation workflows ~\cite{Lu_2024}.

Across domains, the evidence shows that EAI4S addresses the binding constraint of limited research capacity. EAI4S systems extend experimental duty cycles far beyond human-only workflows, which are typically limited to a few experiments per operator per day. This shift from human-limited execution to continuously operating EAI4S infrastructure multiplies effective research capacity per scientist and makes EAI4S a practical necessity where workforce scale is the dominant constraint.

While these results demonstrate the general productivity advantages of EAI4S, their practical significance is best understood through concrete, end-to-end laboratory workflows.

\section{Case Study: Water Quality Analysis Experiments Empowered by EAI4S}
\label{sec:case}

To illustrate how these productivity gains materialize in routine laboratory practice, we present a case study of a water quality analysis experiment empowered by EAI4S, conducted at the National Institute of Clean-and-Low-Carbon Energy in China.

A standard water quality analysis experiment follows a tightly sequenced workflow spanning instrument calibration, chemical handling, sample preparation, and waste processing. Although the analytical objective is routine, successful execution depends on a long chain of precise, safety-critical physical operations, where an error at any stage can invalidate the entire result. As summarized in Table~\ref{tab:operators_mapping}, the five experimental stages decompose into 22 distinct atomic task operators, invoked 34 times across the workflow, highlighting the operational density and fragility of human-executed laboratory procedures.

The procedure begins with electrode preparation and calibration, including rinsing and alcohol cleaning, opening and closing protective caps, inserting electrodes into instruments, and performing calibration and pH verification, often followed by HCl calibration. These steps require careful handling to avoid contamination, drift, or sensor damage. Throughout the experiment, operators must interact with heterogeneous instrument interfaces, including physical buttons and touchscreens, while repeatedly positioning electrodes, cups, and beakers with high spatial accuracy. Liquid handling is a dominant source of operational risk, involving pipetting, pouring, and transferring fluids among beakers, test tubes, and sample containers, where small spillage or dosing errors can compromise results. The experiment concludes with waste liquid disposal and instrument reset, where improper handling can create safety risks or cross-contamination. Collectively, this sequence illustrates that throughput is limited by the cumulative complexity of coordinated manual actions rather than by analytical difficulty.

In current laboratory practice, water quality analysis experiments rely heavily on well-trained operators, typically holding undergraduate or graduate degrees in chemistry, chemical engineering, or related fields. Each of the five stages requires approximately 5 minutes of careful manual execution, resulting in roughly 25--30 minutes of hands-on time per experiment after accounting for transitions and documentation. As a consequence, a single operator typically completes only 12 to 15 experiments per day under normal working conditions. This limitation is driven by repetitive, safety-critical manual operations such as electrode handling, liquid transfer, and instrument interaction. Even with skilled personnel, these workflows remain susceptible to spillage, contamination, or procedural deviations, reinforcing that human capacity is the dominant constraint on experimental throughput.

\begin{table}[t]
\centering
\caption{Atomic Operator Decomposition of a Water Quality Analysis Experiment}
\label{tab:operators_mapping}
\renewcommand{\arraystretch}{1.25}
\begin{tabular}{p{4.2cm} p{8.2cm}}
\toprule
\textbf{Experiment Step} & \textbf{Required Atomic Operators} \\
\midrule

Electrode Rinsing &
Pick up electrode; open electrode cap; rinse electrode with liquid; pour liquid; wipe electrode with alcohol; close electrode cap; place electrode \\

Electrode Calibration &
Open electrode cap; insert electrode into instrument; press physical button; operate touchscreen interface; wait for calibration completion; remove electrode; close electrode cap \\

Calibration Verification &
Insert electrode into instrument; place cup or beaker into instrument; press physical button; operate touchscreen interface; read pH value; remove electrode; remove cup or beaker \\

HCl Calibration &
Open reagent bottle; operate pipette; dispense HCl into container; insert electrode into instrument; press physical button; operate touchscreen interface; remove electrode; close reagent bottle \\

Waste Handling &
Grasp beaker or container; pour waste liquid; operate valve (close); dispose waste container; clean electrode with alcohol \\

\bottomrule
\end{tabular}
\end{table}

\begin{figure}
\centering
\includegraphics[width=0.8\columnwidth]{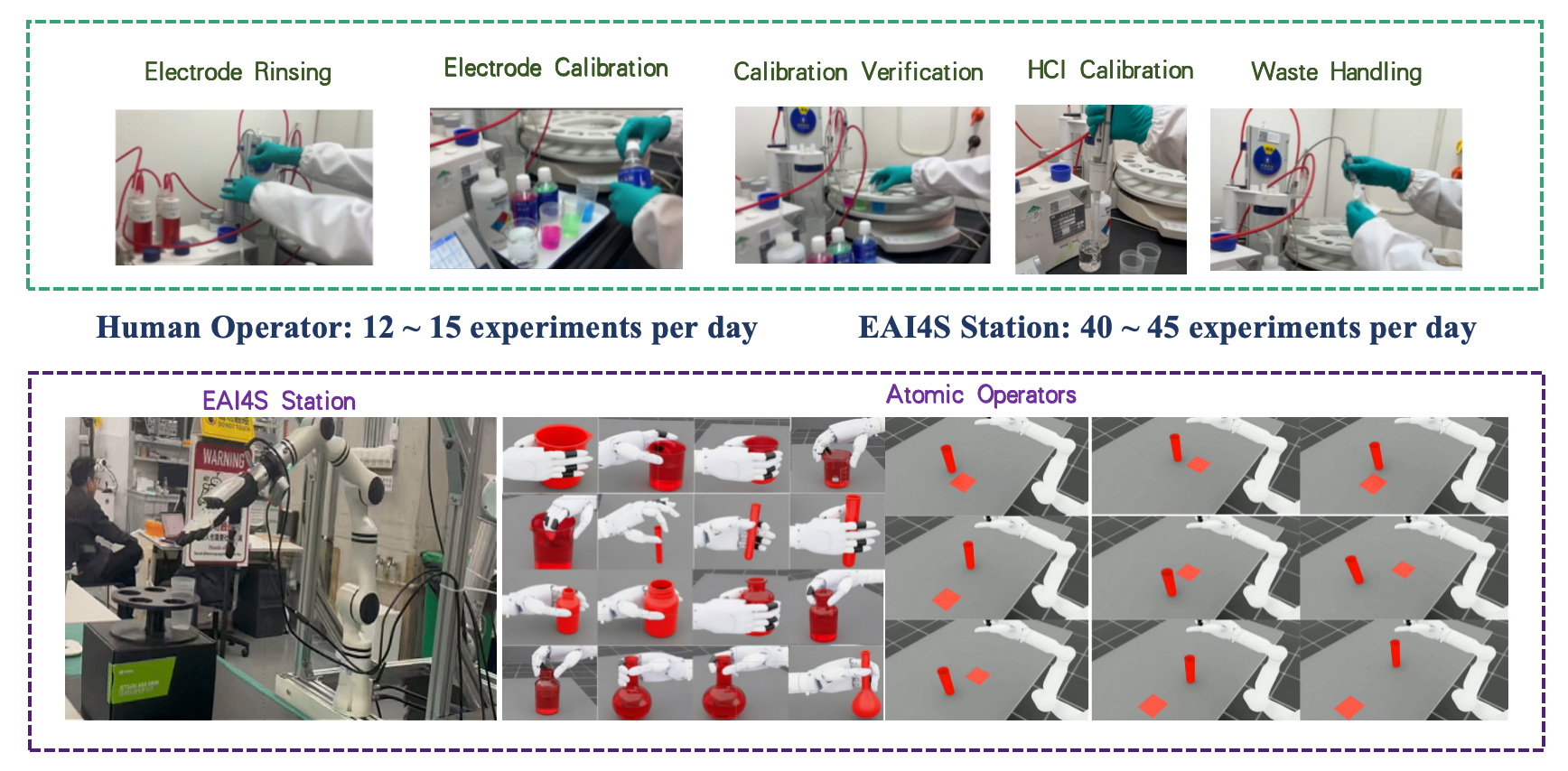}
\caption{EAI4S Multiplies Experimental Throughput in Water Quality Analysis}
\vspace{10pt}
\label{fig:throughput}
\end{figure}

As illustrated in Fig.~\ref{fig:throughput}, the EAI4S station is an integrated embodied AI platform centered on a robotic arm with a dexterous robotic hand, designed to execute laboratory experiments through sequences of atomic task operators. The multi-joint arm provides accurate positioning and reach across the workspace, while the dexterous hand supports grasping, rotation, insertion, and tool use, enabling reliable interaction with laboratory instruments, containers, and samples. Operating at approximately human-equivalent motion speed but without fatigue, the station can sustain extended duty cycles. In our current deployment, this yields a throughput of roughly 40--45 experiments per day, compared to 12--15 experiments per day for a skilled human operator, implying that one EAI4S station delivers the effective capacity of approximately three human operators while maintaining repeatability and procedural consistency.

This case study provides concrete evidence that EAI4S addresses the primary bottleneck in scientific workflows, limited human execution capacity, by transforming routine experiments into scalable, continuously operable processes. The implications of this capacity multiplication extend beyond individual laboratories and directly bear on the structural constraints faced by research systems in the Global South.

\section{EAI4S as a Necessity for the Global South}
\label{sec:necessity}

The water quality case study reflects a broader pattern observed at the system level: the scientific output gap between advanced economies and Global South countries is driven primarily by differences in research capacity rather than per-researcher productivity.

\begin{figure}
\centering
\includegraphics[width=\columnwidth]{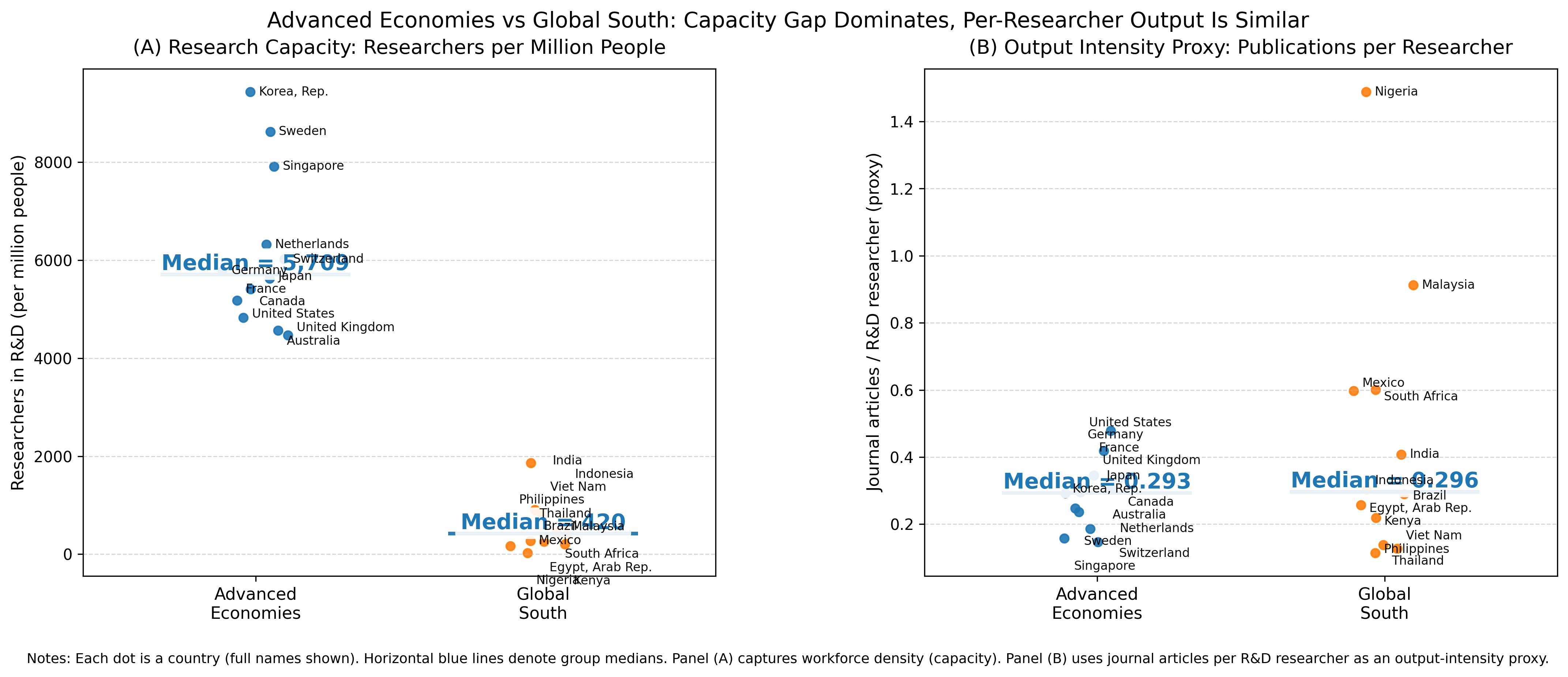}
\caption{Researcher Density Dominates Differences in Scientific Output}
\vspace{10pt}
\label{fig:comparison}
\end{figure}

The scientific output gap between advanced economies and Global South countries is often interpreted as a difference in research quality or capability. As illustrated in Fig. ~\ref{fig:comparison}, our compiled indicators support a more operational diagnosis. The dominant constraint is research capacity, measured as the number of researchers available per capita, rather than a systematic shortfall in per-researcher output intensity.

Research capacity is quantified using the World Bank World Development Indicators series SP.POP.SCIE.RD.P6, which reports researchers engaged in research and development per million people. This indicator captures the density of the R\&D workforce under standard definitions used in cross-country statistical reporting ~\cite{researchers}.

In our sample, the median researcher density in advanced economies (5709 researchers per million people) is more than an order of magnitude higher than that of Global South countries (420 researchers per million people), indicating a structural difference in the scale of available scientific labor. This pattern is consistent with broader global assessments that show research resources and researcher pools are concentrated in a small set of economies ~\cite{UNESCO_report}.

To evaluate output intensity, we use the World Bank WDI series IP.JRN.ARTC.SC, which counts scientific and technical journal articles indexed in Scopus across science and engineering fields, following the National Center for Science and Engineering Statistics taxonomy ~\cite{pubs}. 

We compute an output-intensity proxy as the number of scientific and technical journal articles divided by the estimated number of researchers. Using this measure, the median publications per researcher are similar across advanced economies and Global South countries in our dataset. This observation is consistent with global science assessments showing that, once normalized by researcher counts, differences in publication intensity across countries are substantially smaller than differences in total output driven by workforce size ~\cite{output}. 

Independent cross-country bibliometric analysis also finds that developed economies outperform developing-country groups on most input and impact dimensions, but not on publications per researcher, indicating that workforce scale is a dominant driver of aggregate output gaps ~\cite{GonzalezBrambila_2016}. This interpretation should nonetheless be treated cautiously, as international co-authorship, publication attribution practices, and incomplete researcher statistics can distort country-level ratios, particularly for smaller or data-sparse research systems. Nevertheless, the combined evidence cleanly distinguishes two levers: increasing the number of effective scientific hands and increasing the output per hand. The first lever dominates in magnitude.

This capacity gap has direct implications for scientific workflow design. When the researcher base is small, researchers must absorb a disproportionate burden of technical tasks such as sample handling, instrument operation, monitoring, data capture, and repeated experimentation. These tasks scale poorly with limited personnel and reduce the throughput of discovery even when scientific expertise is comparable. Conventional remedies, such as expanding the research workforce and laboratory technician base, require long time horizons and sustained investment.

EAI4S directly addresses the primary structural constraint underlying the scientific output gap: limited research capacity.  Prior work has demonstrated that autonomous scientific systems can sustain higher experimental throughput and reproducibility than manual workflows, even when expert supervision is limited ~\cite{king2009automation, burger2020mobile}. In capacity-constrained research environments, such systems effectively multiply the productive reach of individual scientists.

From a Global South perspective, EAI4S should be treated as enabling infrastructure rather than a premium capability. Open science and open-source AI already reduce barriers to accessing methods and models, but they do not eliminate the execution gap created by limited personnel and fragile laboratory operations. UNESCO’s open science recommendations highlight the importance of reducing barriers to participation and reuse, but effective participation ultimately depends on the ability to run experiments, collect data, and maintain reliable workflows ~\cite{UNESCO_report}. By converting computational intelligence into physical action and standardized procedures, EAI4S increases the effective research capacity per scientist and can therefore contribute directly to narrowing the scientific output gap driven by workforce scale.

As a result, for Global South research systems where scaling the human scientific workforce is structurally slow, EAI4S should be treated as essential capacity-building infrastructure rather than an optional advanced technology.

\section{Infrastructure Needs for EAI4S}
\label{sec:infra}

The preceding section establishes that the core scientific gap facing the Global South is limited research capacity driven by workforce scale. If EAI4S is to function as a capacity-multiplying infrastructure rather than a niche automation tool, its deployment must be supported by the physical and digital systems that enable continuous experimentation.

In this sense, EAI4S is fundamentally an infrastructure problem. While access to models and algorithms is increasingly global, the ability to sustain embodied experimentation depends on networks, computation, storage, robotics, and standardized operational frameworks. This section evaluates EAI4S readiness by identifying the key foundations required for scalable adoption in low-resource research environments.

\textbf{1) Network infrastructure:}
EAI4S primarily relies on local-area networking for real-time robot control, instrument coordination, and experiment logging, with wide-area connectivity used intermittently for synchronization, software updates, and remote oversight. Per station, peak network demand during execution is modest: low-resolution monitoring video (e.g., 720p, compressed) along with peripheral logs typically requires 5Mbps. At laboratory scale, this demand aggregates linearly. A deployment of 10–20 EAI4S stations requires a stable LAN capacity on the order of 100 Mbps, a level well within modern Ethernet capabilities but not always available in legacy laboratory infrastructure.

Wide-area connectivity presents a more binding constraint in the Global South. According to the ITU, median mobile broadband download speeds in low-income countries are typically below 15 Mbps \cite{ITU_FF2023}. Fixed broadband penetration and reliability are often lower in public research institutions outside major urban centers. As a result, continuous cloud-dependent operation is impractical. EAI4S systems must therefore adopt a local-first networking model, relying on high-quality LANs for execution and scheduling bandwidth-efficient, asynchronous backhaul for periodic data and log transfer rather than sustained high-throughput connectivity.

\textbf{2) Storage infrastructure:}
EAI4S increases experimental duty cycle and data volume, making storage a binding constraint. A single EAI4S station typically generates 10 GB per day from video streams, sensor logs, and experiment metadata. At small laboratory scale (e.g., 10~15 stations), this translates to roughly 5 TB of new data per month.

To function under intermittent connectivity, EAI4S requires tiered local storage: fast storage for run-time logs, experiment artifacts and intermediate analysis, and bulk storage for long-term retention. In practice, this implies a minimum on-site capacity on the order of 50 TB for continuous operation.

By contrast, UNESCO assessments identify insufficient local storage and curation capacity as a key barrier preventing researchers in low-resource settings from converting experimental data into durable scientific assets \cite{UNESCO_OpenScience}. Without adequate storage infrastructure, increased experimental autonomy cannot translate into sustained scientific output or reproducibility \cite{Wilkinson2016FAIR}. 

In our first-hand engagement with Egyptian partner laboratories, experimental data are often stored and backed up at the individual level rather than through institutionally managed systems, making long-term curation and reproducibility fragile.

\textbf{3) Compute infrastructure:}

EAI4S requires predictable on-site compute for perception, state estimation, safety monitoring, and manipulation planning. A practical sizing metric is sustained edge inference throughput per station. In our setup, a single-station setup with 2 cameras and real-time detection/segmentation plus grasping and safety checks typically requires on the order of \emph{50 TOPS} of sustained inference. 

What limits adoption in many Global South labs is not the feasibility of edge compute, but the availability of reliable local hardware and support. A cross-field survey on computing for research in Africa reports that a majority of respondents rate local hardware resources and computing power as insufficient, and that local infrastructure constraints are a recurrent blocker \cite{Rahal2022AfricaComputing}. Consequently, EAI4S deployments should provision station-level accelerators as dedicated infrastructure, rather than assuming shared clusters or ad hoc workstations can sustain continuous embodied experimentation. 

Our experience in Egyptian labs shows that while centralized or shared computing resources exist, reliable always-on compute at the laboratory floor is limited, constraining real-time embodied experimentation and motivating dedicated station-level edge compute.

\textbf{4) Robotic infrastructure:}
Robotic hardware remains the most visible capital barrier for EAI4S adoption. Commercial collaborative robot arms are typically priced in the tens of thousands of USD before accounting for end-effectors, sensors, integration, and safety systems. The EAI4S station used in our case study is built around a modular robotic arm and dexterous hand with a total hardware cost of approximately \emph{USD 10,000}, lowering the entry barrier by an order of magnitude.

Despite this reduction, most laboratories in the Global South today operate with little to no access to robotic manipulation platforms. Experimental workflows therefore remain almost entirely human-executed, even for routine and repetitive procedures. This absence of robotic infrastructure, rather than a lack of algorithms or models, is a primary reason why EAI4S has not yet translated into practical capacity gains in low-resource research environments. 

\textbf{5) Software and procedural infrastructure:}
Open-source software stacks and open-science platforms play a critical role in leveling access to methods, models, and analytical tools, reducing entry barriers for Global South institutions \cite{UNESCO_OpenScience}. However, access alone is insufficient. In the absence of shared EAI4S software abstractions and standardized experimental procedures, EAI4S experiments remain difficult to align, reproduce, and compare across laboratories. Practical EAI4S deployment therefore requires not only open and maintainable software for orchestration, device control, calibration, and logging, but also procedure-level standardization that makes experimental execution machine-interpretable and portable. Without explicit definitions of operator primitives, protocol parameters, instrument states, and provenance metadata, results generated by different EAI4S systems cannot be reliably reproduced, even when underlying algorithms are shared. Extending FAIR principles from data to EAI4S procedures is thus essential: scientific outputs must be accompanied by executable, standardized protocols that enable experiments to be rerun under comparable conditions across sites and over time \cite{Wilkinson2016FAIR}.

\begin{figure}
\centering
\includegraphics[width=0.8\columnwidth]{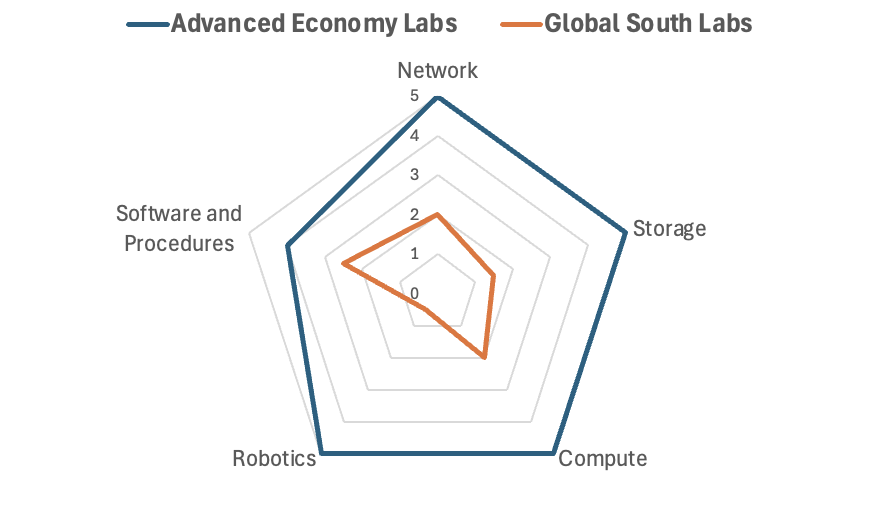}
\caption{Readiness Analysis of Global South EAI4S Adoption}
\vspace{10pt}
\label{fig:readiness}
\end{figure}

\paragraph{Readiness Assessment.}
Figure~\ref{fig:readiness} reports readiness scores computed using the threshold-based scheme defined in this section. Advanced-economy laboratories score near the maximum because they typically meet or exceed the minimum operational requirements for multi-station EAI4S deployment, including $\sim$100,Mbps local networking, tens of terabytes of on-site storage, dedicated edge accelerators for continuous inference, and routine access to general-purpose robotic manipulators. In contrast, Global South laboratories score lower because local infrastructure commonly falls below these thresholds: robotic manipulation platforms are rare, sustained edge compute and local storage are limited, and networking is often insufficient or costly for continuous operation. The resulting score differences therefore reflect measurable infrastructure gaps relative to defined EAI4S operational requirements, not differences in scientific expertise or algorithmic access.

\section{From Policy to Action: Enabling EAI4S at Scale}
\label{sec:policy}

The analysis in this paper identifies a clear economic inflection point for large-scale adoption of EAI4S. When the total system cost of an EAI4S station, including robotics, compute, storage, networking, maintenance, and integration, falls below the annual cost of employing a single trained researcher, EAI4S shifts from a niche automation tool to a rational, capacity-multiplying investment. Crucially, this cost threshold is shaped not only by hardware prices but also by the presence of shared software stacks and global standards, which reduce integration effort, duplication, and long-term operational overhead. In Global South contexts, where scientific capacity is constrained by workforce scarcity and slow training pipelines, lowering both system and coordination costs defines a scalable pathway for EAI4S to narrow the global scientific capacity gap.

\textbf{Recommendation 1: Invest in infrastructure where EAI4S crosses the cost parity threshold.}
Governments, funding agencies, and research institutions in the Global South should prioritize targeted investments in local infrastructure, particularly networking, storage, and edge computing, once EAI4S system costs approach or undercut annual researcher costs. At this point, each deployed station can sustainably multiply experimental throughput without increasing dependence on scarce human labor, enabling continuous experimentation even under constraints in connectivity and manpower.

\textbf{Recommendation 2: Drive down EAI4S system costs through research and engineering.}
The responsibility is reciprocal. EAI4S researchers and system builders must treat cost reduction as a first-class research objective, alongside autonomy and performance. This includes designing hardware-efficient models, modular robotic platforms, low-bandwidth operation modes, and open, reusable software stacks. Lowering system cost directly expands the set of institutions for which EAI4S becomes economically viable, accelerating global diffusion.

\textbf{Recommendation 3: Establish global EAI4S standards for reproducible embodied science.}
Cost alone is insufficient without alignment. The absence of global standards for EAI4S procedures, operator primitives, data schemas, and provenance makes experimental results difficult to compare or reproduce across sites. An international standardization effort—extending FAIR principles from data to embodied experimental protocols—is essential to ensure that EAI4S enables replicable, cumulative science rather than isolated automation.


\bibliographystyle{ACM-Reference-Format}
\bibliography{ref}
\end{document}